# Semiconductor-compatible topological digital alloys

Yunfan Liang[1], *Damien West[1], Shunda Chen[2], Jifeng Liu[3], Tianshu Li[2], Shengbai Zhang[1]

[1]Rensselaer Polytechnic Institute, Troy, NY 12180, USA

[2]Department of Civil and Environmental Engineering, George Washington University, Washington, DC 20052, United States

[3]Thayer School of Engineering, Dartmouth College, 14 Engineering Drive, Hanover, New Hampshire 03755, USA

*Corresponding author.

## Abstract

Recently, GeSn alloys have attracted much interest for direct-gap infrared photonics and as potential topological materials which are compatible with the semiconductor industry. However, for photonics, the high-Sn content required leads to low detectivity, associated with poor material quality, and the (>35%) Sn required for topological properties have been out of reach experimentally. Here, we demonstrate that by patterning the Sn distribution within Ge, the electronic properties have a far greater tunability than is possible with the random alloy. For the GeSn $\delta$-digital alloy (DA) formed by confining Sn atoms in atomic layer(s) along the [111] direction of Ge, we show that ~10% Sn can lead to a triple-point semimetal. These findings are understood in terms of Sn ordering causing spatial separation of Sn and Ge band edges, leading to band inversion. This mechanism can also lead to a weak topological insulator, Weyl semimetal, and enables tunable direct bandgaps down to 2 meV, covering the entire infrared range. This DA induced topological properties are also identified in compound semiconductors, such as $InAs_{1-x}Sb_x$, showing the general applicability of the DA design for realizing topological properties on conventional semiconductor platforms. Our findings not only point to a new class of currently unexplored topological systems accessible by epitaxy, but also establish the promise of low-Sn GeSn DAs for application as infrared laser diodes and photodetectors in Si photonic integrated circuits and infrared image sensors.

## Introduction

Since the discovery of the quantum Hall effect, the study of topological materials has become an important branch of modern condensed matter physics. These include topological insulators, Dirac semimetals, Weyl semimetals, and, more recently, triple-point semimetals., including topological insulators [1, 2, 3], Dirac semimetal [4, 5, 6], Weyl semimetals[7, 8, 9], and more recently, triple point semimetals [10, 11, 12, 13], which exhibit nontrivial bulk band topology and robust surface states. Potential applications such as topological field-effect transistors [14, 15], spintronics devices [16, 17], topological superconductors [18], and topological quantum computing[19] have also been proposed. Despite the many promises, actual technological applications still face many challenges, including the growth of stable high-quality samples and integration with the existing semiconductor platforms. For example, even though previous computational modeling has extensively screened nearly 40,000 compounds and identified ~8,000 potential topological material candidates [20], none of the representative candidates is compatible with existing semiconductor platform, thereby greatly limiting the potentials of quantum device fabrication and scalable integration for practical applications.

In this regard, group IV materials and their alloys can be a natural choice for overcoming these challenges, as they have good compatibility with Si and Ge based technologies. Recent theoretical predictions suggest that incorporation of ~10% Pb [21] or 25-28% Sn [22, 23] into Ge should cause a topological phase transition. While promising, such high metal concentrations pose experimental challenges due to their large lattice mismatch. Pb incorporation is currently limited to below 5.2% [24] and experimental efforts increasing the Sn concentration up to 35% [25, 26, 27, 28] have failed to show the existence of such a topological transition, suggesting the true transition occurs at even higher concentrations[29]. Alternatively, this could be due to poor quality of the material, as growth of GeSn with such high Sn concentrations can result in both the formation of extended defects [30, 31] and Sn segregation [31, 32]. In fact, such issues already appear at lower concentrations, e.g., GeSn Si-based photonic applications require >18% Sn to cover infrared sensing in the 3-5 μm atmospheric window and >25% Sn for the 8-10 μm window [33, 34]. So far, these high Sn GeSn infrared photodetectors are suffering from low detectivity resulting from such materials issues [35, 36].

For such GePb and GeSn random alloys, the band gap naturally decreases with increased metal content. Straightforwardly, at large enough concentrations, this can lead to a small gap which may be associated with a topological phase transition. However, the practicality of incorporating such high metal concentrations and the associated poor material quality, motivates the exploration of alternative physical mechanisms for band gap reduction. Instead of simple alloying, here we propose spatially separating the valence and conduction bands, by varying alloy composition, to allow for unprecedented control of the band structure of the resulting alloy. In particular, if the valence and conduction bands are associated with different types of atoms, as shown schematically in Fig. 1, the polarization dipoles which develop at the interface between these regions can dramatically reduce (or even close) the band gap. This reminds us of a digital alloy (DA), where the concentration in different regions alternate as a step function [37]. As control of atomic layer composition has been demonstrated for group IV system epitaxy [38], such systems may provide a practical way to incorporate quantum materials in today's Si based devices. Note that the mechanism presented in Fig. 1 is by no means limited to group-IV. It can also work for compound semiconductors (as will be discussed below).

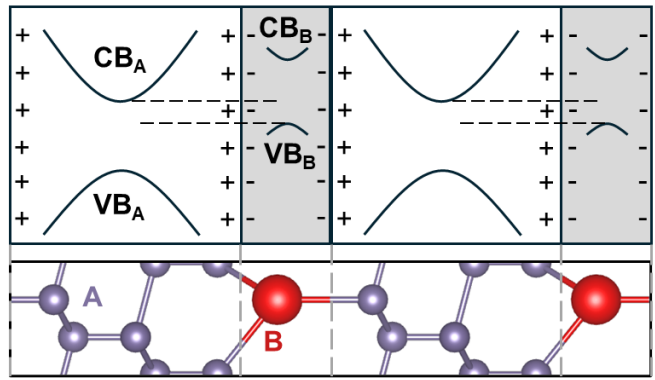

Figure 1 Schematic illustration of a digital alloy containing layers of atoms of type A and B. Atomic relaxation at the interface results in an interfacial dipole causing a rigid shift in the relative band structures, leading to an overall band gap smaller than gaps of the isolated systems.

In this work, we investigate the $Ge_{1-x}Sn_x$ DAs with an Sn concentration x in the range of 8 to 16%, based on density-functional theory (DFT) calculations, combined with the Wannier function based tight - binding modeling. In our design of the DA, the Sn is incorporated as $\delta$-doping layers along the [111] direction of diamond cubic Ge. The resulting DAs exhibit non-trivial topological properties at a minimum Sn concentration of only 10%, far lower than Sn concentrations which have already been incorporated in $Ge_{1-x}Sn_x$ alloys [28]. Further, the DAs are highly compatible with the current semiconductor technology, as the lattice mismatch with Ge in the lateral directions can be as small as 1.2%, compared to 4% lattice mismatch for epitaxial Ge-on-Si photonic devices widely used in Si PICs [39]. In this range of Sn, the DA is found to transition from a small-gap normal insulator to either a weak topological insulator or a triple-point semimetal. Furthermore, by applying an external magnetic field to break time reversal and mirror symmetry, we also obtain the Weyl semimetal phase. Using the same design principle, we also show DAs can lead to topological phase transitions in $InAs_{1-x}Sb_x$ DA with x ~ 16-33%. As such, not only are DAs of GeSn a powerful platform for developing monolithic Si-compatible quantum/photonic devices, but the principle of embedding periodic lower-dimensional structures in 3D materials is a fruitful direction for design of new classes of materials with rich 3D topological properties.

## Results

### 1Sn/5Ge DA as a prototypical triple-point semimetal

The DA is defined as a Ge crystal with Sn atoms doped into a single atomic layer in [111] direction of face centered cubic (FCC) cell. Take the configuration shown in Fig. 2(a), where one layer of Ge is completely replaced by Sn in every 6 atomic layers, denoted as the 1Sn/5Ge configuration, as a prototype. The total Sn concentration for this configuration is ~17%. The DA has a hexagonal cell with the basis vectors **a, b** and **c** along $[1\bar{1}0]$ $[10\bar{1}]$ and $[111]$ of FCC cell, respectively. The corresponding first Brillouin zone is shown in Fig. 2 (b). The DA belongs to the $C_{3v}$ point group with 1 rotational axis and 3 mirror planes all parallel to the **c**-direction. The calculated band structure, including spin-orbit coupling (SOC), is shown in Fig. 2 (c) with atomic orbital projection on s and p orbitals of Ge. In contrast to bulk Ge, where the s* (s anti-bonding) derived conduction band minimum (CBM) is approximately 0.8 eV higher than the p derived valence band maximum (VBM) at $\Gamma$, for the DA, band inversion takes place with s* being 0.16 eV lower than p derived band. This band inversion results in non-trivial band topology which is confirmed by calculation of the $Z_2$ topological invariant[1, 40] which was found to be 1 for the $k_z = 0$ plane and 0 for the $k_z = \frac{\pi}{c}$ plane. Along the Γ-A path, all k points have C$_{3v}$ symmetry, so the bands can be labeled by the irreducible representations of the C$_{3v}$ double group [41], whose character table can be found in supplementary material (see Table 1 in SM). As shown in the zoomed in plot, band crossing occurs between the double degenerate s* band belonging to $D_{1/2}$ and two single p bands belong to $D'$ and $D''$ irreducible representations, respectively. Since these three bands belong to different irreducible representations, these two triple degenerate crossing points centered at ($k_z = 0.128 \frac{\pi}{c}$) and separated by ~0.3meV are protected by crystalline symmetry. As such, the DA can be classified as a triple-point (TP) semimetal [11]. The 3D band plotting for plane A ($k_z = 0.128 \frac{\pi}{c}$) shows linear band dispersion near the TP, which confirms the 3D semimetal nature of this 1Sn/5Ge DA. The TPs appear in pairs located around $k_z = \pm 0.128 \frac{\pi}{c}$, respectively which are connected by time-reversal symmetry as shown in Fig. 2 (e).

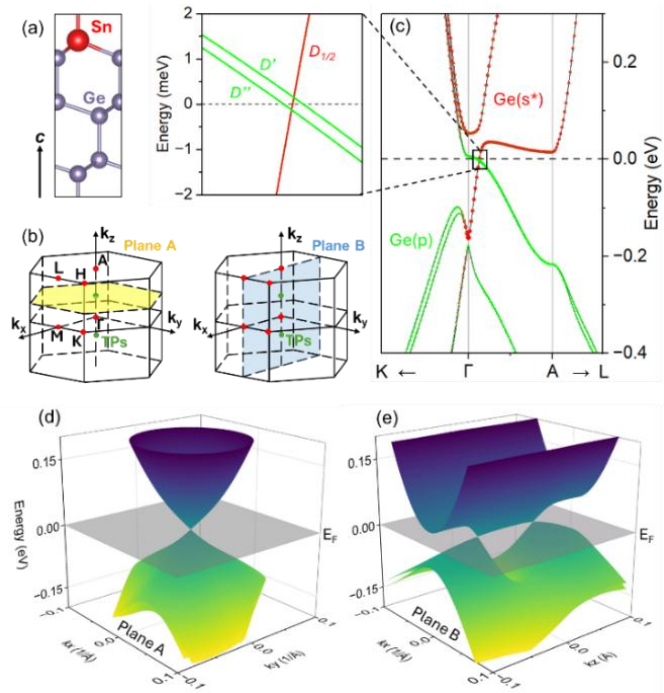

Figure 2 **Triple-point semimetal phase in 1Sn/5Ge DA:** (a) Atomic structure and (b) first Brillouin zone for the 1Ge/5Sn DA configuration. (c) The corresponding band structure and a zoomed-in plot near the TP. (d) and (e) are the 3D band structures for states near the Fermi level, for $k_z = 0.128\frac{\pi}{c}$ [Plane A in (b)] and $k_y = 0$ [Plane B in (b)]. The grey plane in (d) and (e) represent the fermi level.

The Dirac point of a topological semimetal can be considered as two copies of Weyl points with opposite Chern number [40]. A TP semimetal is similar to the Dirac semimetal[11], and the TPs (at $\vec{k}_{TPs}$) are connected to the TPs at $-\vec{k}_{TPs}$ through a fermi-arc[13]. To investigate the surface properties, we choose a slab with surface parallel to **a** and **c** axis, such that the TPs do not fold onto each other and the 2D surface Brillouin zone reduces to plane B (folded along $k_y$) shown in Fig. 2(b). The simulated surface spectrum is shown Fig. 3 (a) where it can be seen that two surface states emerge from the nearly degenerate triple points (labeled as TPs in the figure) and cross at $\bar{\Gamma}$. Note that this degeneracy at $\bar{\Gamma}$ is protected by time reversal symmetry and $E(\vec{k}) = E(-\vec{k})$, so along the $\bar{k}_z$ direction these surface states emerge from TPs (at $\vec{k}_{TPs}$) and vanish at the other pair of triple points at $-\vec{k}_{TPs}$. In the $\bar{k}_x$ direction, the surface states merge into the conduction and valence bands. Fig. 3(b) shows the surface fermi-arc and surface spin texture at the fermi-level. Since the splitting between the two TPs is very small ~0.3meV, the fermi-arc looks similar to that in the Dirac semimetal where different sides of the fermi-arc have opposite spin direction and the direction of the spin varies when go through each side of the arc [42]. Since the point $\vec{k}$ and $-\vec{k}$ have opposite spin, the backscattering induced by non-magnetic defects are eliminated due to strong cancellation [43].

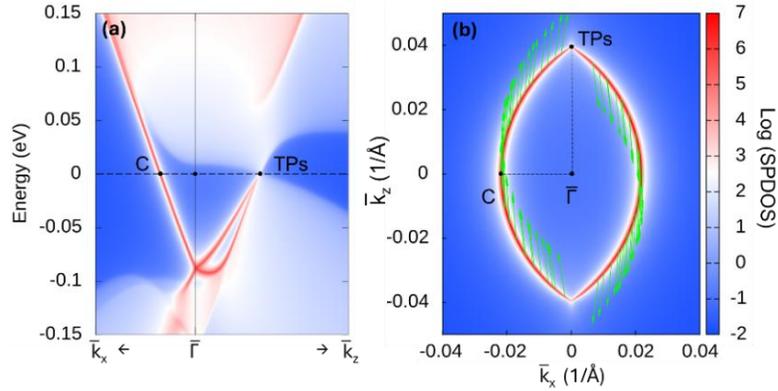

Figure 3 **Topological surface states:** (a) Simulated surface states of the $[11\bar{2}]$ surface for angle resolved photoemission spectroscopy (ARPES) and (b) the corresponding surface Fermi-arc and spin texture (green arrows). Here the 2D BZ for surface is formed by folding the bulk BZ in $k_y$ direction, so the $\bar{k}_x$ and $\bar{k}_z$ in the plot are associated with the $k_x$ and $k_z$ of bulk. The color represents the logarithm of projected surface density of state (PSDOS). The red color denotes the states localized to the surface, the white color denotes the bulk states and blue denotes no states. The simulations are carried out by WannierTools [44]

## Physical origin of band topology in DAs

The non-trivial band topology in 1Sn/5Ge DA arises from the band inversion near Γ, which can be treated as a negative band gap (-0.16eV). However, for the random alloy at the same Sn concentration as DA (16.7%), a positive gap was found of approximately 0.25eV (see Fig. S1 in SM), consistent with previous calculations [29]. This indicates that even with fixed Sn concentration, the gap can be significantly reduced just by altering the distribution of Sn atoms. To understand the physics behind this, we take the 1Sn/5Ge DA and special quasi-random structure (SQS) [45] to represent the random alloy (RA), as prototypes to investigate the effect of the Sn distribution within a 54-atoms supercell. As the lattice constants of the DA and random alloy were found to be nearly identical (~0.1% difference in the lateral direction and 0.5% difference in the vertical direction) and have negligible effect on the band structure, to facilitate the subsequent analysis we use the lattice constant of the fully relaxed DA for all calculations.

To understand the effect of the Sn distribution on the electronic structure, we first investigate the band structure of the DA and RA in the absence of atomic relaxation – wherein atoms are placed at the ideal positions of a Ge lattice strained to match the lattice constant of the relaxed DA, as shown in Fig. 4(a,b). Here, the band gap of the DA (0.13 eV) is found to be substantially smaller than the band gap of the RA (0.36 eV), indicating that the Sn distribution of DA can significantly reduce the band gap even without atomic relaxation. Two more SQS configurations are tested as detailed in Fig. S1 of SM, yielding similar results. These results can be primarily understood due to the degree of localization of the CBM of the DA and RA. For the DA, the CBM is localized to the Ge region, as shown in the inset of Fig. 4 (c), and hence is associated with Ge s* states. On the other hand, for the RA, the CBM becomes delocalized and hence contains contributions from both the Ge s* states and the higher energy Sn s* states, leading to an overall increase in the energy. When considering the evolution of the band structure from the DA to the RA, the VBM is relatively unchanged (slight decrease) while the CBM monotonically increases (see 'Evolution from DA to RA' section in SM) in energy as the CBM has an increasing contribution from Sn. This understanding is shown schematically by the energy level diagram in Fig. 4 (c).

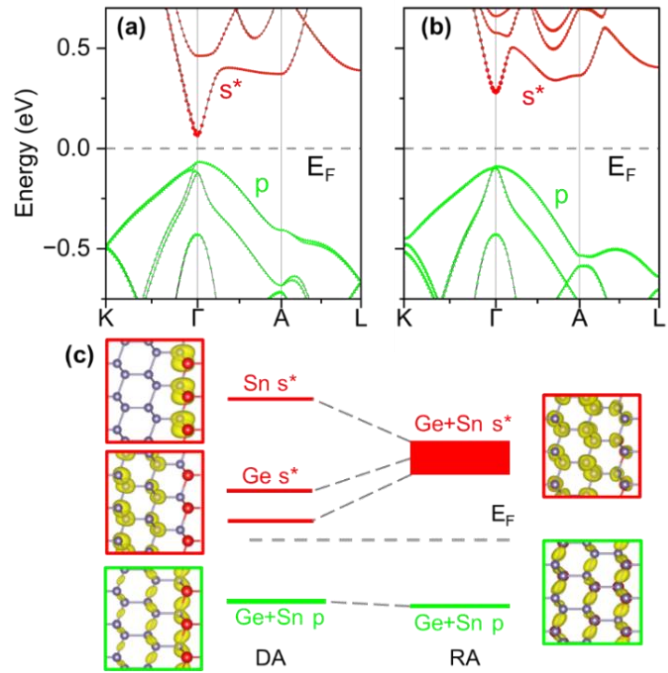

Figure 4 **Effect of Sn distribution without relaxation:** Band structures for (a) unrelaxed DA, (b) unrelaxed RA, all in the lattice constant of fully relaxed DA. The average electrostatic potential energy is used to align the band structures. The fermi level is marked by horizontal dash lines in each plot. (c) sketches the energy levels at Γ for DA and RA near fermi-level, where the thick bar represents a group of energy levels. The inserts show the charge distribution (yellow cloud) for corresponding level with red and purple balls representing Sn and Ge atoms, respectively. The two Ge s* in DA have similar electron distribution, so only the CBM is shown in insert. For the thick bar in RA, only the lowest level (CBM) is shown in insert.

For the unrelaxed DA, the band order near the Fermi level is still the same as bulk Ge, so the lattice distortion induced by Sn atoms also plays an important role in causing band inversion. For ease of description, we will refer to chemical bonds parallel to the c-direction as "vertical bonds" and other chemical bonds as "in-plane bonds". After relaxation, the Ge-Ge and Ge-Sn bond length are relaxed, on average, from 2.50Å to 2.48Å and 2.58Å, respectively. Upon atomic relaxation, the strain induced by the longer Ge-Sn bonds is released by pushing the adjacent Ge atoms away along the c-direction, reducing the in-plane bond angles ($\beta' < \beta$) and increasing the angle between in-plane and vertical bonds ($\alpha' > \alpha$) as sketched in Fig. 5 (a).

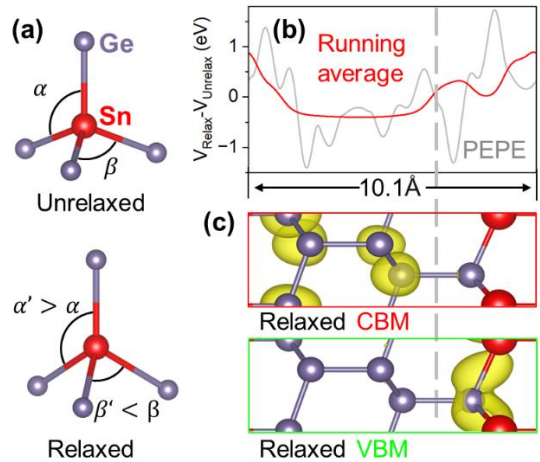

Figure 5 **Effect of relaxation:** Panel (a) sketches the distortion around the Sn atoms. Panels. (b) shows the change in planar-averaged electrostatic potential energy (PEPE) (grey) and the change in running average of PEPE (red) after relaxation. The interval width of 3.35Å and 3.23Å are applied to calculate the running average of PEPE in unrelaxed and relaxed lattice, respectively. To avoid the overall translation during relaxation the position of Sn atomic layer is aligned to calculate the difference. (c) are the CBM and VBM electron distribution (yellow cloud) at Γ for the relaxed DA

As Sn pushes neighboring Ge atoms away, an electrostatic dipole layer develops on each side of the Sn. This can be seen in Fig. 5(b), where relaxation causes the electron potential energy to increase at the Sn layer and decrease among the Ge layers. This further reduces the gap as the CBM is localized to the Ge layers (with decreasing energy) while the VBM becomes localized to the Sn layer (with increasing energy), as shown in Fig. 5(c). Electronically, the DA can then be seen as a periodic quasi-2D structure embedded in bulk Ge wherein the VBM of the 2D structure is higher in energy than the CBM of the bulk 3D structure, resulting in a topological phase transition where the gap is reduced from 0.13 eV to -0.16 eV. In contrast, for the case of the RA, no such topological phase transition takes place. Not only is the unrelaxed band gap of the RA larger (0.36 eV), but the relaxation associated with the Sn distribution results in randomly oriented dipoles, instead of forming a quasi-2D structure. As a result, the band gap reduction due to relaxation in the RA is only 0.11 eV (resulting in a positive gap of 0.25 eV), substantially smaller than the 0.29 eV reduction in the DA.

Note that the bandgap reduction achieved by DA does not rely on any special properties of group-IV elements. As a result, this design strategy can be readily extended to other alloy systems. One example in the III–V family is the InAs$_{1-x}$Sb$_x$ DA. According to previous experimental studies [46], InAs, InSb, and their alloys are all semiconductors. However, 2(InAs)/1(InSb) and 5(InAs)/1(InSb) DAs exhibit triple-point semimetal phase, as shown in Fig. S3 in SM.

## Band topology dependence on Sn concentration and DA configuration

Besides the 1Sn/5Ge configuration discussed in Fig. 2-5, topological properties can also be observed in other DA configurations. For the DA, the configuration can relate to both the spacing between Sn doping layers as well as the concentration of Sn within the Sn doping layer. To investigate the dependence on configuration, (1) we investigate how altering the Sn concentration within the Sn doping layer x, within our prototypical DA system, i.e. 1(Sn$_x$Ge$_{1-x}$)/5Ge, affects the topological properties. Then we consider the effects of altering the spacing, by both (2) investigating equally spaced Sn layers of spacing n, i.e. 1Sn/nGe, and systems which have (3) two distinct spacings $n_1$ and $n_2$ (let $n_1 \leq n_2$), i.e. 1Sn/$n_1$Ge/1Sn/$n_2$Ge. Finally, we investigate the applicability of these configurations to as-grown materials by considering (4) the effects of being unable to confine Sn to a single atomic layer, but instead have Sn distributed within bilayers that include in-plane bonds, i.e. 2(Sn$_{0.5}$Ge$_{0.5}$)/4Ge.

(1) The Sn concentration for each configuration can be adjusted by varying the density of Sn atoms within the doping layer. First the separation in the **c**-direction is fixed at 5 Ge layers, we investigated lowering the Sn concentration by replacing either 1 or 2 of the Sn atoms in a $3 \times 3$ Sn doping layer by Ge, yielding Sn concentrations x of 89% and 78% within such atomic layer, respectively. These configurations were chosen to maintain the C$_{3v}$ symmetry of the underlying lattice which would remain unbroken in a random 1(Sn$_x$Ge$_{1-x}$)/5Ge DA. The band structures of 1(Sn$_x$Ge$_{1-x}$)/5Ge, with decreasing doping concentration, are shown in Fig. 6 (a-c). Here we see that as the Sn concentration is reduced, the lowest s* state increases in energy relative to the highest p state (band gaps of -0.16, -0.06, and 0.08 eV, respectively) leading to a vanishing of the band inversion when the Sn concentration reaches 78% wherein the system becomes a normal insulator (NI).

(2) Next, we fix the Sn concentration to 100% but adjust the separation in the **c**-direction to get different configurations. As shown in Fig. 6(a, d-f), as the separation along the **c**-direction is increased from 5 to 11 Ge layers, which reduces the total Sn concentration from 16% to 8.3%, the gap becomes less negative with the highest p state (green) decreasing in energy relative to the lowest s* state (red). Additionally, increasing the spacing reduces coupling between highest p states, which are confined to the quasi-2D structure, resulting in a notable reduction in the $\Gamma - A$ dispersion. Further, as these configurations belong to the $C_{3v}$ point group, if band inversion exists at $\Gamma$ a symmetry-protected TP also exists along the Γ-A path. As can be seen, increasing the spacing between Sn layers leads to a topological phase transition. When the layer spacing is between 5Ge layers and 9 Ge layers, each configuration exhibits non-trivial topology, resulting in the emergence of a TP semimetal. As the layer spacing is increased to 11 Ge layers, the gap is no longer inverted (having a very small $E_g = +2$ meV) resulting in a normal insulator. This tiny direct bandgap is in sharp contrast to the GeSn RA of the same composition (8.3% Sn), which has a bandgap ~0.55 eV [47]. This indicates that DAs can exhibit non-trivial topology with a Sn concentration near 10%, which is significantly lower than the >35% predicted in traditional GeSn alloys [28, 29]. Although we have focused on the inverted band regime, the tunability of the bandgap at low Sn concentration (<10%) suggests this could be a promising platform for THz/IR detection, with the gap potentially spanning from semimetal to that of bulk Ge. For example, at the same low Sn composition of 8.3%, one could potentially achieve perfectly lattice-matched GeSn heterostructures and QWs by engineering the atomic ordering of DAs vs. random alloys towards high-performance infrared lasers and photodetectors on Si platform.

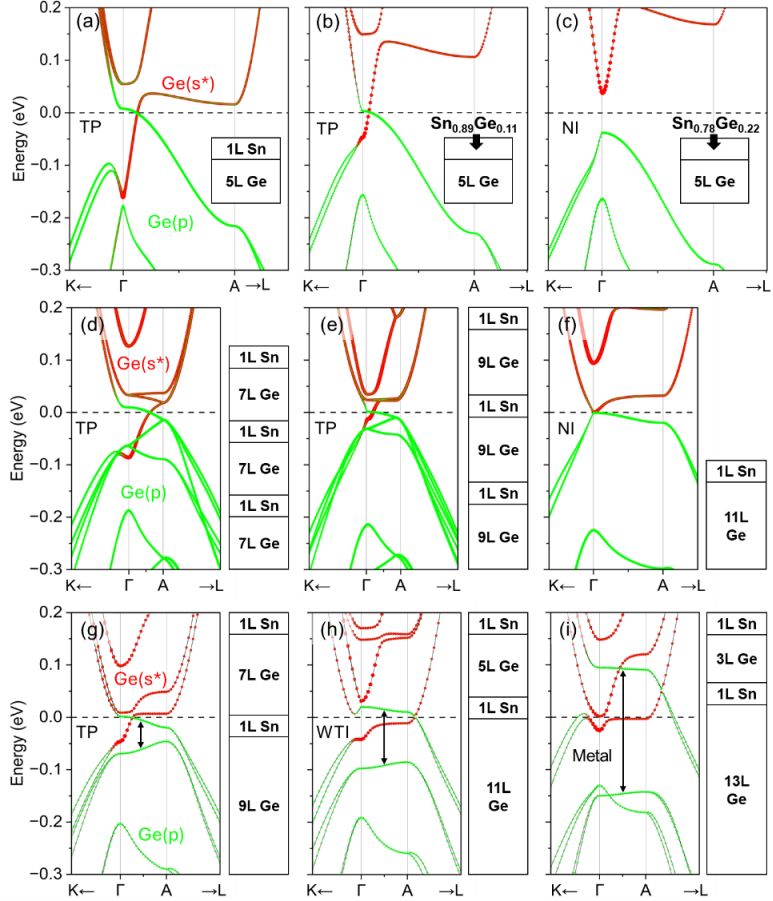

Figure 6 **Various topological phases in GeSn DA:** The band structure of 1($Sn_xGe_{1-x}$)/5Ge DA with (a) x=1, (b) x=0.89 and (c) x=0.78 as illustrated by the sketches inside the band structures. The corresponding total Sn concentrations are (a) 16.7, (b) 14.8 and (c) 12.9%, respectively. (d), (e) and (f) are the band structure for 1Sn/nGe DA with (d) n=7, (e) n=9 and (f) n=11, as illustrated by the sketches next to the band structures. The corresponding total Sn concentrations are (d) 12.5, (b) 10 and (c) 8.3%, respectively. For the diamond structure the repeating unit along [111] is 6 atomic layers, so the minimum supercell for (b) the 1 Sn/7 Ge case contains 24 layers and for (c) the 1 Sn/9 Ge case, it contains 30 layers. (g), (h) and (i) show band structure for 1Sn/$n_1$Ge/1Sn/$n_2$Ge DA with 18 layers of atoms per-period ($n_1$+ $n_2$=16). The smaller separation between 2 adjacent Sn layers is (g) $n_1$=7, (h) $n_1$=5 and (i) $n_1$=3 layers of Ge respectively as illustrated by the sketches next to the band structures. Letters TP, WTI and NI denote, respectively, triple-point semimetal, weak topological insulator and normal insulator. The fermi level is set to zero in all panels.

(3) The distribution of Sn layers can also be ununiform along **c**, providing greater freedom to engineer the band structure. We take the supercell with 18 layers of atoms with two layers fully replaced by Sn as a prototype. As sketched in Fig. 6 (g-i), the number of Ge layers between the Sn layers is larger ($n_2$) on one side but smaller on the other side ($n_1$). Since the total number of atomic layers in one period is fixed, the configurations can be defined by $n_1$ as $n_2 = 16 - n_1$. Note that in this series of calculations, the Sn content remains unchanged leading to a minimal change in lattice constant (< 0.01 %). Here the difference can be traced to the interaction between the highest and second highest p states, the splitting of which is indicated by the arrows in Figs. 6(g-i). Both of these p states are localized to the quasi-2D structures which are no longer connected by translational symmetry, and hence are no longer degenerate at A. As the minimum distance between Sn layers is reduced the interaction between these states increases, leading to larger splitting. While the TP semimetal is clearly seen for 1Sn/7Ge/1Sn/9Ge, when $n_1$ is reduced in the 1Sn/5Ge/1Sn/13Ge configuration the increasing of the highest p state causes the crossing between s* and p to disappear along the $\Gamma - A$ path, with a small gap opening near A. As there is no longer a crossing along the $\Gamma - A$ line, this indicates the topological properties of the material have changed, with the band gap becoming inverted at both $\Gamma$ and $A$. Calculation of the $Z_2$ invariants as 0;(001) reveal the 1Sn/5Ge/1Sn/11Ge configuration to be a weak topological insulator (WTI). Further reducing $n_1$ yielding the 1Sn/3Ge/1Sn/13Ge configuration, the splitting between p-states is further increased to the degree that the anticrossing between lowest s* and highest p state occurs above the second lowest s* state, resulting in changing occupation and a transition to a metallic state. A quantitative analysis of how $n_1$ and $n_2$ modify the splitting and dispersion of the p-states along Γ-A, within the context of a tight-binding model, can be found in the 'Tight-binding model' section in SM.

(4) Although so far, we have considered Sn doping confined to isolated atomic layers, the observed non-trivial band topology does not require such perfect ordering. This is important as controlling the composition of individual atomic planes during growth along the [111] direction is exceedingly difficult. As the [111] surface with three dangling bonds per atom is much more reactive than the surface terminated at vertical bonds with a single dangling bond per atom, controlling the Sn content per Ge bilayer is much more feasible. We investigated this by taking a 3×3×1 supercell of our prototypical 1Sn/5Ge DA, where instead of restricting our Sn to a single atomic layer they are randomly distributed throughout the bilayer, forming 2($Sn_{0.5}Ge_{0.5}$)/4Ge DA.

While some degree of disorder may be expected during the growth process, we find that the predicted topological properties are robust against variation. The band gap at Γ from SQS (red dash) agrees well with the average value of the 10 randomly generated configurations and the gaps of all configurations of the bilayer DA, 2($Sn_{0.5}Ge_{0.5}$)/4Ge, are found to be more negative than the 1Sn/5Ge DA, as shown in Fig. S6 in SM, indicating that the disorder of Sn within the bilayer tends to enhance band inversion. The reason for this enhances band inversion can be understood that all Sn atoms are still confined in the quasi-2D structure, so the s* state (CBM) that avoiding these layers almost unaffected, but the p (VBM) is raised up by the disorder in the chemical environment within the quasi-2D structure, now containing Ge-Sn, Sn-Sn, and Ge-Ge bonds. Such effect of atomic disorder has been discussed in our previous publication [21].

**Tuning DA properties via magnetic field and Weyl semimetal phase**

The TP semimetal can be considered as an intermediate state between Dirac and Weyl semimetals, where breaking inversion symmetry causes the Dirac semimetal to become a TP semimetal and further reducing the symmetry of the TP semimetal (either through breaking time-reversal symmetry or $\sigma_v$ mirror symmetry) will lead to a Weyl semimetal. In principle, any magnetic field in the **c**-direction will break both, leading to the formation of a Weyl semimetal. Figure 7 shows the resulting Weyl semimetal which forms by explicitly including a Zeeman coupling $B_z \mu_B g$ of $0.02 eV$ in our Hamiltonian. As the g-factors are quite large for topological semimetals, with $g = 20$ for $Na_3Bi$[48] and $g = 40$ for $Cd_3As_2$[49], and for short period superlattices, $g = 105$ for InAsSb/InSb [50], this Zeeman splitting corresponds to a reasonable magnetic field strength, on the order of 3~20T. Here we note that the 0.02 eV Zeeman splitting is only for reference, with better experimental resolution the $B_z$ can be substantially reduced. With such a magnetic field, the $C_{3v}$ point group symmetry reduces to $C_3$. The doubly degenerate s* bands split into two single bands, so each TP splits into two Weyl points. As shown in Fig. 7 (a), there are 8 Weyl points labeled by $W_{1,2,3,4}$ and $W'_{1,2,3,4}$, which are protected by the $C_3$ symmetry. The Chern number for $W_{1,3}$ and $W'_{2,4}$ are calculated to be +1, while the Chern number for $W_{2,4}$ and $W'_{1,3}$ are calculated to be -1, which shows the Chiral anomalous nature of the Weyl semimetal.

The surface states of a semi-infinite slab system along $[11\bar{2}]$ direction have been simulated. With respect to bulk Ge, the "top" and "bottom" surface are identical, however for the DA, the bottom surface contains exposed Sn while on the top, all Sn are covered, as shown in Fig. 7 (b). Two surface states connecting to the $W_1$, $W'_1$ and $W_4$, $W'_4$ can be observed. The Fermi arc of the top and bottom surfaces taken for $E_F = E_1$ are shown in Fig. 7 (c) and exhibit the characteristic half loop on the top surface connecting the degenerate Weyl points of opposite Chern number $W_1$ and $W'_1$, with the other half loop on the bottom surface. Similarly shown in Fig. 7 (d), for $E_F = E_2$, the Fermi arcs connect $W_4$ and $W'_4$, forming a complete loop only when both top and bottom surfaces are considered. As is understood, for carrier flow through the arcs, backscattering is effectively suppressed, as the other side of each arc is on the opposite surface, where their electronic states are fully isolated by the semi-infinite slab. Although in theory $W_2, W'_2$ and $W_3, W'_3$ should give you another set of Fermi arcs, they are not discernible in our calculation as in the case of the Dirac semimetal $Na_3Bi$ under external magnetic field[51].

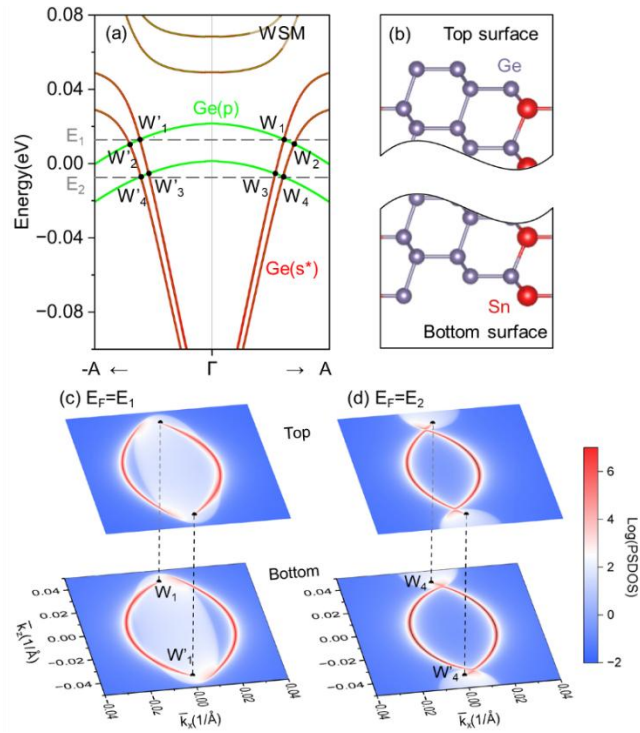

Figure 7 **Weyl semimetal phase:** (a) The band structure near Γ with the Zeeman field along **c**-direction. (b) shows the surface applied for the simulation of surface fermi arc. (c) and (d) show the fermi-arc for 'top' and 'bottom' surface at $E_1$ and $E_2$, respectively. The black dots mark the position of Weyl points at the selected energy, which are (c) $W_1$, $W'_1$ at $E_1$ and (d) $W_4$, $W'_4$ at $E_2$. The color in (c) and (d) represents the logarithm of projected surface density of state (PSDOS). The red color denotes the states localized to the surface, the white color denotes bulk states and blue denotes no state. The simulations are carried out by WannierTools [44]

## Conclusion

In this work first-principles calculation reveals that the construction of a DA, in which Sn mono/bilayers are embedded in bulk Ge to form quasi-2D structures, can have a profound effect on the band structure and topological properties of the GeSn alloy. Through systematic calculations wherein the spacing between Sn layers and Sn concentration within dopant layers are varied, we show that with Sn concentrations as little as 10%, the resulting material can transition between a normal insulator, a topologically non-trivial triple point semimetal, a weak topological insulator, and with the inclusion of a magnetic field, a Weyl semimetal. Further, these topological properties were found to be robust, with Sn disorder within a bilayer actually enhancing the band inversion. While these properties are intimately tied to the 2D structures embedded in bulk, it is important to note we refer to 3D band topology, confirmed by the 3D cone structure, $Z_2$ invariant, and surface states. As such, this makes these DA systems intrinsically different from the 2D band topology in the type-III aligned heterostructure [52] and 2D materials, like graphene, germanene or stanene [53]. This understanding can be extended to include other systems, such as creating 2D networks of 1D nanowires or 3D networks for 0D nano dots in host materials, which provides a new direction for designing topological materials.

In terms of experimental implementation, the lattice mismatch between the designed low-Sn DAs and Ge is only 1.2%, much less than the 4% lattice mismatch of epitaxial Ge-on-Si devices widely used in silicon PICs. Moreover, as discussed earlier, the band topology is maintained even if Sn and Ge intermix in the bilayers. Therefore, epitaxial growth of GeSn DAs on existing Ge/Si virtual substrates is feasible, providing a new class of silicon-compatible topological materials. Furthermore, low-Sn small bandgap GeSn DAs (e.g. similar to Fig. 6(f)) can be applied to infrared laser diodes and photodetectors in Si PICs and infrared image sensors, addressing two grand challenges in Si-based topological electronics and infrared photonics simultaneously.

## Method

### Structure optimization and band structure calculations

The Vienna Ab initio Simulation Package (VASP) [54] based on the projector augmented wave method is applied for first principal calculations. Structural relaxation of the alloy system is performed using the local-density approximation (LDA) functional [55], which yields accurate lattice constants for Ge and α-Sn bulk (within ~0.2% of their experimental values). The band structures presented and their associated wavefunctions used for calculations of the topological invariants are determined using the META-GGA functional with modified Becke-Johnson (MBJ) potential [56, 57] wherein the c-mbj parameter is set to

1.2 to reproduce the band gap of bulk Ge and the semimetal nature of α-Sn. The validity of the META-GGA calculated band structure in alloy system is further confirmed by comparing with hybrid functional (HSE06) [58] results as detailed in Fig. S7 in SM. A 15×15×$N_z$ Monkhorst-Pack k-points grid was used, where $N_z$ was chosen such that $N_z$ times the number of atomic layers is approximately equal to 42 with $N_z$=7 for 1Sn/5Ge (6 layers), $N_z$=3 for 1Sn/11Ge (12 layers), $N_z$=2 for 1Sn/7Ge (24 layers) and all 1Sn/$n_1$Ge/1Sn/$n_2$Ge configurations (18 layers), and $N_z$=1 for 1Sn/9Ge (30 layers). For the calculations of SQS for RA, 1($Sn_xGe_{1-x}$)/5Ge DA and 2($Sn_{0.5}Ge_{0.5}$)/4Ge DA, 54-atoms supercells constructed by repeating the 1Sn/5Ge DA cell 3 times in each lateral direction are applied. The supercell has lateral dimensions about three times that of 1Sn/5Ge DA with similar vertical dimension, so a 5x5x7 Monkhorst-Pack k-points grid was chosen. An energy cutoff of 300 eV, combined with the convergence criteria of $10^{-7}$eV and $10^{-3}$ eV/ Å for electronic and ionic relaxations, respectively.

## Topological properties calculations

The topological properties, including $Z_2$ invariant, surface states, surface fermi-arc, spin textures and Chern number of each Weyl point are calculated by WannierTools [44] with the Hamiltonian under the maximally localized Wannier basis fitted by Wannier90 library[59]. The s and p orbitals for all atoms are applied as the basis Wannier functions to fit the band structures from first-principal calculations.

## Data availability

Data are available from the corresponding author upon reasonable request.

## Code availability

The codes used in this work can be purchased at VASP https://www.vasp.at/, downloaded at Wannier90 http://www.wannier.org/, and downloaded at WannierTools https://www.wanniertools.org/

## Acknowledgements

We would like to thank Yonghang Zhang for bringing our attention to the promise of digital alloys. We would also like to thank Jin Hu and Shui-Qing Yu for fruitful discussions.

This work was supported by μ-ATOMS, an Energy Frontier Research Center funded by the U.S. Department of Energy, Office of Science, Basic Energy Sciences under award DE-SC0023412. We also acknowledge supercomputer time provided by NERSC under DOE Contract No. DEAC02-05CH11231, time provided on TAMU FASTER and Purdue Anvil made available by ACCESS through allocation TG-DMR180114, and the CCI at RPI.

## CRediT authorship contribution statement

**Shengbai Zhang:** Conceptualization, Project administration, Supervision, Writing – review & editing. **Yunfan Liang:** Investigation, Formal analysis, Writing – original draft. **Damien West**: Project administration, Supervision, Writing – original draft. **Jifeng Liu:** Conceptualization, Writing – review & editing. **Shunda Chen:** Investigation, Writing – review & editing. **Tianshu Li:** Resources, Writing – review & editing.

## Competing interests

The authors declare no competing financial interests.

# Supplementary Materials for "Semiconductor-compatible topological digital alloys"

Yunfan Liang[1], *Damien West[1], Shunda Chen[2], Jifeng Liu[3], Tianshu Li[2], Shengbai Zhang[1]

[1]Rensselaer Polytechnic Institute, Troy, NY 12180, USA

[2]Department of Civil and Environmental Engineering, George Washington University, Washington, DC 20052, United States

[3]Thayer School of Engineering, Dartmouth College, 14 Engineering Drive, Hanover, New Hampshire 03755, USA

*Corresponding author.

Table S1. Character table of C₃ᵥ double group, where A1, A2, and E are from the typical single group representation and $D_{1/2}$, D', and D'' are new representations in the double group.

| $C_{3v}$ | $E$ | $R$ | $2C_3$ | $2RC_3$ | $3\sigma_v$ | $3R\sigma_v$ |
|---|---|---|---|---|---|---|
| $A_1$ | 1 | 1 | 1 | 1 | 1 | 1 |
| $A_2$ | 1 | 1 | 1 | 1 | −1 | −1 |
| $E$ | 2 | 2 | −1 | −1 | 0 | 0 |
| $D_{1/2}$ | 2 | −2 | 1 | −1 | 0 | 0 |
| $D'$ | 1 | −1 | −1 | 1 | $i$ | $-i$ |
| $D''$ | 1 | −1 | −1 | 1 | $-i$ | $i$ |

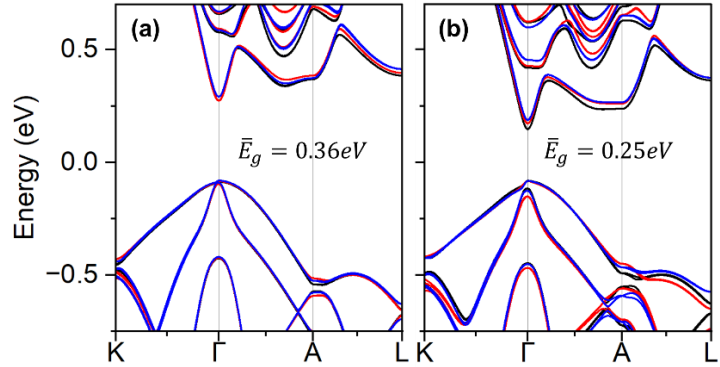

Figure S8 The SQS band structures for (a) unrelaxed and (b) fully relaxed Se₀.₁₇Ge₀.₈₃ random alloy. The three different colors (black, red, and blue) in each panel correspond to different randomly chosen SQS configurations. The average band gap, $\bar{E}_g$, for the three unrelaxed SQS configurations is 0.36eV. Atomic relaxation lowers this gap by 0.11eV, leading to a 0.25eV average band gap. For both relaxed and unrelaxed cases, the different SQS configurations have similar band structures, the results presented in the main text as for the blue configuration.

## Evolution from DA to RA

To investigate the evolution from DA to RA, we first construct a 3×3×1 supercell of the 1Sn/5Ge DA. We then remove Sn atoms from the GeSn bilayer and randomly distribute them in the Ge region. As the number of Sn atoms which are moved increases, the system becomes more like a RA. We denote the fraction of RA, $x$, as $C_{Sn}^{GR}/C_{Sn}$, where $C_{Sn}^{GR}$ is the concentration of Sn in the Ge region (GR) and $C_{Sn}$ is the Sn concentration in the supercell. As shown Fig. S2(a-d), with the RA percentage increasing from 0% to 100%, the electron at the CBM becomes more distributed around Sn atoms and less localized, which indicates that the CBM has an increasing contribution from higher energy Sn s* states. As a result, the energy of the CBM monotonically increases as shown in Fig. S2 (e), which supports the understanding in Fig. 4 (c) of the main text. Note that the CBM is already quite delocalized at 66% and has substantial contribution from the Sn s* state and further increasing to 100% RA has a significantly smaller effect on the CBM energy. The VBM is relatively unchanged in all configurations (difference <30meV), so the gap increase is mainly due to the increase of the CBM.

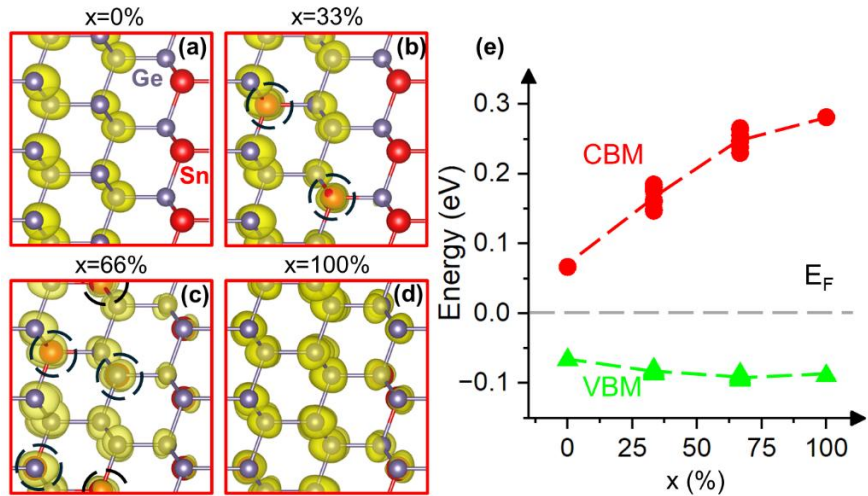

Figure S9 (a-d) shows the CBM electron distribution for the unrelaxed $Sn_{0.17}Ge_{0.83}$ with RA percentages, x, of (a) 0%, (b) 33%, (c) 66%, and (d) 100%. The dashed circles mark Sn atoms in the Ge region. (e) shows the corresponding energies of the VBM and CBM where average electrostatic potential energy was used as a reference for band alignment. For 33% and 66% case 10 randomly generated configurations were tested for each case to account for alloy randomness and the configuration whose band gap is closest to the average value is chosen as prototype to show CBM electron distribution in (b) and (c), respectively.

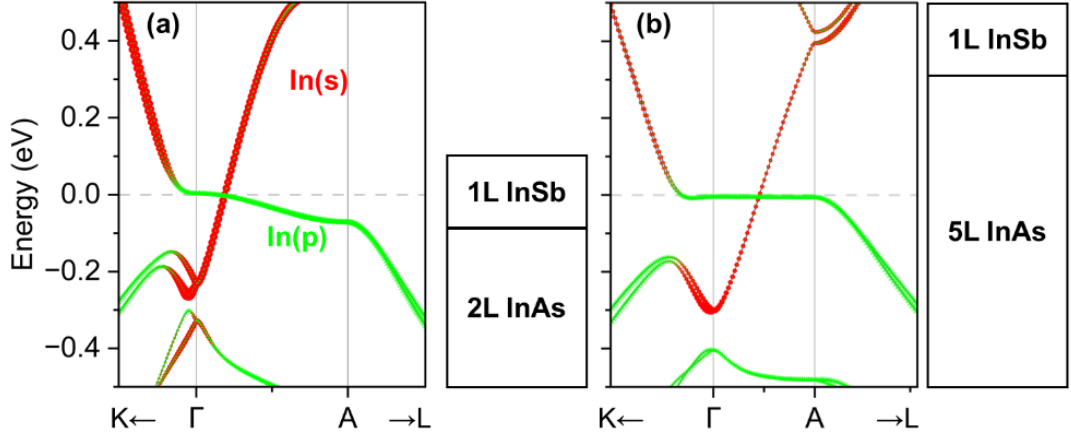

Figure S10 The band structures of (a) 2(InAs)/1(InSb) DA and (b) 5(InAs)/1(InSb) DAs with orbital projected onto p orbital (green) and s orbital (red) of Indium atoms. The symmetry of III-V DAs is the same as the Ge/Sn DA, so they are both triple-point semimetals.

**Tight-binding model.**

In Fig. 6 (g-i), along Γ-A path the highest and second highest p bands are both including 2 single bands belonging to D' and D'' representations with very small splitting(~0.3meV). All other bands nearby belong to $D_{1/2}$ representation. Since coupling can only happen between bands with the same irreducible representations, we only need to consider the coupling between two bands with D' representation and between two bands with D'' representation. These two sets of p bands almost overlap with each other (~0.3meV offset), so we only present the results for the two bands belonging to the D' representation.

As the p bands are confined to the quai-2D structure containing Sn, we investigate how well the splitting of these p bands can be represented by a simple tight-binding model in which the hopping parameters $t_1(n_1)$ and $t_2(n_2)$ depend on the separation between Sn layers by $n_1$ or $n_2$ Ge atomic layers, as shown in Fig. S3 (a). In this case, the Hamiltonian can be written as,

$$H = \begin{pmatrix} E_0 & t_1 + t_2 e^{-ik_z L} \\ t_1 + t_2 e^{ik_z L} & E_0 \end{pmatrix} \quad (S1)$$

where $L$ is the length of supercell in the **c** direction and $E_0$ is the energy of the p bands before interaction with respect to the Fermi level. Diagonalizing Eq. (S1) yields the eigenvalues,

$$E(k_z) = E_0 \pm \sqrt{t_1^2 + t_2^2 + 2t_1 t_2 \cos(k_z L)}. \qquad (S2)$$

As shown in Fig. S3(b-d), Eq. (S2) describes well the DFT band structure along the $\Gamma \to A$ path with the fitted parameters shown in Table. S1. As the distance between Sn layers increases, the associated hopping parameter between layers naturally decreases and it is found that both $t_1$ and $t_2$ are simply the interlayer hopping parameter $t(n)$ for the particular distances $n = n_1$ and $n_2$. Plotting the fitted hopping parameter $t(n)$ in Fig. S4, $t$ is found to decay towards 0 exponentially with a $t = -(395\ meV)e^{-n/2.50}$.

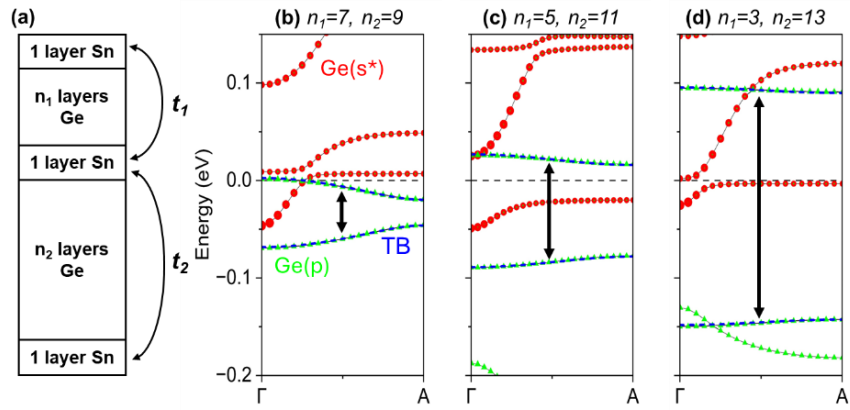

Figure S11 (a) The sketch of the DA configurations and the hopping parameters. (b-d) Show the band structure from DFT calculation for each configuration along Γ-A path, with the orbital projection on Ge s* and p orbitals. The blue dash line shows the band structure from the tight binding Hamiltonian for the p bands near fermi-level, whose splitting is marked by black arrows. The fermi level is set to 0.

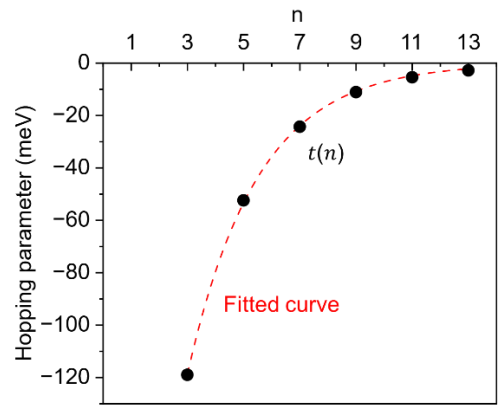

Figure S12 The hopping parameter t(n) with different n (black dots) and the fitted exponential curve (red dash).

Table S2. The fitted parameters for the tight-binding Hamiltonian.

| Configurations | $E_0$ (meV) | $t_1$ (meV) | $t_2$ (meV) |
|---|---|---|---|
| $n_1=7, n_2=9$ | -33.0 | -24.3 | -11.1 |
| $n_1=5, n_2=11$ | -31.1 | -52.4 | -5.4 |
| $n_1=3, n_2=13$ | -26.4 | -119.0 | -2.78 |

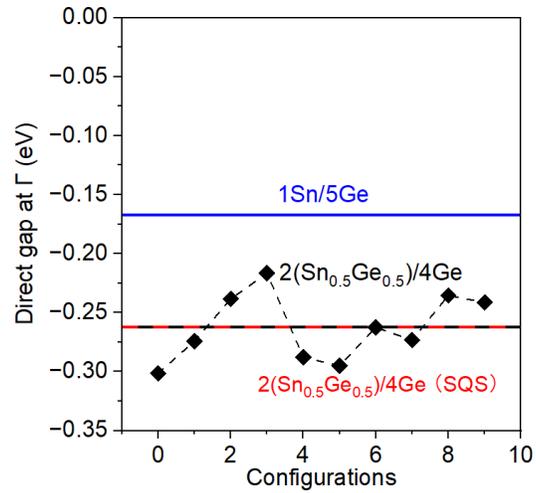

Figure S13 The direct band gap at Γ of 1Sn/5Ge and 2(Sn0.5Ge0.5)/4Ge DA. The blue line represents the results for 1Sn/5Ge DA. The black dots are the results for 10 randomly generated 2(Sn0.5Ge0.5)/4Ge DA with the average marked by black solid line and the red dash line marks the results for SQS of 2(Sn0.5Ge0.5)/4Ge DA.

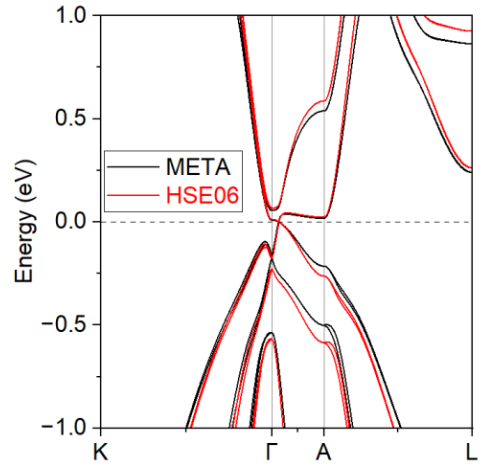

Figure S7 Band structure for 1Sn/5Ge DA calculated by META-GGA functional(black) and HSE06 functional (red). Both methods show the semimetal nature of the DA with similar depth of band inversion at Γ indicating that the essential physics explored in this work are well described by the META-GGA functional